\documentclass[manuscript]{aastex}

\usepackage{color}
\usepackage{natbib}

\slugcomment{Accepted by the {\it Astrophysical Journal} on January 13, 2015}

\shorttitle{NGC 1266 Nuclear Molecular Gas}
\shortauthors{Glenn et al.}

\begin{document}

\title{NGC 1266:  Characterization of the Nuclear Molecular Gas in an
  Unusual SB$0$ Galaxy} 

\author{J. Glenn, N. Rangwala, P.R. Maloney, \& J. R. Kamenetzky}
\affil{CASA, 389-UCB, Department of Astrophysical and Planetary
  Sciences, University of Colorado, Boulder, CO, 80309} 
\email{jason.glenn@colorado.edu}

\begin{abstract}

With a substantial nuclear molecular gas reservoir and broad, high-velocity CO molecular line wings previously interpreted as an outflow, NGC 1266 is a rare SB$0$ galaxy.  Previous analyses of interferometry, spectrally resolved low-$J$ CO emission lines, and unresolved high-$J$ emission lines have established basic properties of the molecular gas and the likely presence of an AGN.  Here, new spectrally resolved CO $J = 5 - 4$ to $J = 8 - 7$ lines from {\it Herschel Space Observatory} HIFI observations are combined with ground-based observations and high-$J$ {\it Herschel} SPIRE observations to decompose the nuclear and putative outflow velocity components and to model the molecular gas to quantify its properties.  Details of the modeling and results are described, with comparisons to previous results and exploration of the implications for the gas excitation mechanisms.  Among the findings, like for other galaxies, the nuclear and putative outflow molecular gas are well represented by components that are cool ($T_{nuclear} = 6^{+10}_{-2}$ K and $T_{outflow} \sim 30$ K), comprising bulk of the mass (Log $M_{nuclear}/M_{\odot} = 8.3^{+0.5}_{-0.4}$ and Log $M_{outflow}/M_{\odot} = 7.6^{+0.3}_{-0.3}$), and the minority of the luminosity (Log $L_{nuclear}/L_{\odot} = 5.44^{+0.22}_{-0.18}$ and Log $L_{outflow}/L_{\odot} \sim 6.5$) and warm ($T_{nuclear} = 74^{+130}_{-26}$ K and $T_{outflow} > 100$ K), comprising a minority of the mass (Log $M_{nuclear}/M_{\odot} = 7.3^{+0.5}_{-0.5}$ and Log $M_{outflow}/M_{\odot} \sim 6.3$) but the majority of the luminosity (Log $L_{nuclear}/L_{\odot} = 6.90^{+0.16}_{-0.16}$ and Log $L_{outflow}/L_{\odot} \sim 7.2$).  The outflow has an anomalously high $L_\mathrm{CO}/L_\mathrm{FIR}$ of $1.7 \times 10^{-3}$ and is almost certainly shock excited.   

\end{abstract}


\section{Introduction}

Massive galaxies with unusual properties can yield insight into
short-lived (and perhaps episodic) stages in galaxy evolution. The
relatively nearby galaxy NGC1266 \footnote{$z = 0.0072$, distance =
30.06 Mpc (Temi et al. 2009); M $= -22.93$, half-light radius
20.4\arcsec~(Cappellari et al. 2011);
Log $L_{\mathrm{FIR}} (\mathrm{L_{\odot}})$ $= 10.44$ (Kamenetzky et al.
2014)} has been classified variously as an S$0$, SB$0$, SA, and a LINER
galaxy. It is unusual in that it contains a large quantity of nuclear
molecular gas and a high-velocity molecular line component:  Alatalo et al. (2011)
discovered a $1.1 \times 10^9$ M$_{\odot}$ centrally concentrated (60
pc radius) molecular gas mass and what they interpret as a $2.4 \times 10^7$ M$_{\odot}$
molecular outflow extending some 460 pc in radius, with emission
extending to $\pm 400$ km s$^{-1}$ and an outflow rate of 13
M$_{\odot}$ yr$^{-1}$. The nuclear molecular gas concentration is
potentially an indication of a merger event, although no merger
partners are apparent. Because the mass outflow rate exceeds the star
formation rate inferred from the far-IR luminosity, Alatalo et
al. (2011) surmised that an AGN powers the molecular outflow. Although only
weak hard X-ray emission was detected by \emph{Chandra X-ray Observatory}, it is
possible that the AGN is buried beneath Compton-thick absorption.
Nyland et al. (2013) presented VLBA observations indicating a very high
brightness temperature ($> 1.5 \times 10^7$ K) radio continuum
emission 1.2 pc in size, coincident with the densest molecular gas,
and JVLA HI absorption observations consistent with the outflow.
Observations of atomic line emission indicated
velocities reaching $\pm 900$ km s$^{-1}$, the presence of shocked
gas, and a morphology consistent with ``nascent'' radio jets (Davis et
al. 2012). It is possible that the molecular high-velocity gas is an outflow and is associated with the ionized gas, but it could also be in a rapidly rotating
disk.  Integral field spectroscopy and Swift UV/optical
observations indicate central-parsec stellar populations $< 1$ Gyr old
and a post-starburst spectrum (Alatalo et al. 2014).

Recently, Pellegrini et al. (2013) analyzed ground-based and {\it
Herschel Space Observatory} SPIRE observations of CO from $J = 1-0$ to $J =
13-12$ and H$_2$O lines. They concluded that the far-infrared spectrum
was ULIRG-like and, based on comparisons to photon-dominated regions and shock models, that
the gas is likely shock heated, although they could not spectrally resolve the CO lines with SPIRE to
separate the disk and outflow components. Determining the physical
state of the molecular gas is crucial to understanding its excitation
mechanism(s), and determining the masses of the centrally concentrated
gas and the outflow are necessary to determine the lifetime of this
evolutionary phase, how much gas is available for star formation, and
what the fate of the gas will be. Here, we analyze new {\it Herschel} HIFI
observations of CO up to $J = 8-7$ that spectroscopically resolve the
outflow and centrally concentrated disk components, and reanalyze {\it
Herschel} SPIRE FTS observations up to $J = 13 - 12$ in combination
with the HIFI observations to quantify the physical parameters of
these gas components. In \S 2, we present the {\it Herschel} HIFI
and SPIRE observations. In \S 3 we present dust modeling and in \S 4 non-LTE models of the CO
emission that constrain the gas physical conditions.  In \S 5 we
compare to previous results and comment on the possible excitation
mechanisms of the gas. Our conclusions are summarized in \S 6.

\section{Observations and Line Fitting}

Four sets of observations were utilized for this analysis: (1) {\it
Herschel} SPIRE Fourier Transform Spectrometer (FTS) spectroscopy of CO from $J = 4 - 3$ to $J = 13 -
12$ and SPIRE 250 $\mu$m, 350 $\mu$m, and 500 $\mu$m dust continuum photometry, (2) {\it Herschel} HIFI
observations of the $J = 5 - 4$, $6 - 5$, $7 - 6$, and $8 - 7$ lines,
(3) ground-based low$-J$ CO observations from the literature, and (4)
continuum photometry from the literature. The SPIRE FTS (Griffin et al. 2010)
provided continuous spectral coverage from 450 GHz to 1,600 GHz with a
spectral resolution of 1.44 GHz. Bands 1b, 2a, 3a, and 3b of
the HIFI receivers (de Graauw et al. 2010) on board {\it Herschel}
(Pilbratt et al. 2010) were used in single-point Wide Band Spectrometer mode.  
All the {\it Herschel} data were
reduced with HIPE-9.  Both sets of {\it Herschel} observations are
listed in Table 1 and the FTS spectrum is shown in Figure 1.  The dust
continuum emission, with flux density rising with frequency, and
bright CO and H$_2$O water lines are apparent, as are the CI $J = 2-
1$ and [NII] lines. The line fluxes reported in Table 2 were derived
by fitting sinc functions (the intrinsic instrumental line profile, as
the lines are unresolved). The HIFI spectra are shown in Figure 2.
For two of the lines ($J = 5 - 4$ and $J = 6 - 5$) the V and H senses of
polarization were combined. For the other two, only one sense was
included because the baselines were poor for the orthogonal
polarizations.  The ground-based CO lines ($J = 3 - 2$, $2 - 1$, and
$1- 0$) were taken from Alatalo et al. (2011).

Following Alatalo et al. (2011), the CO lines were fit with double
Gaussian functions to account for the broad (outflow) component
(hereafter Broad) and the narrow, central velocity component (hereafter CVC), presumably a disk,
and flat baselines. The FWHM of the CVC was fixed at 123 km/s based on
fits to the $J = 2 - 1$ (FWHM $= 123.9 \pm 1.5$ km s$^{-1}$) and $J =
3 - 2$ lines (FWHM $= 122.2 \pm 1.7$ km s$^{-1}$), which had the
highest S/N. The Broad component line widths were allowed to vary, and
for all but the $J = 1 - 0$ line (for which the Broad component
appears very weak), the line widths are consistent with 278 km s$^{-1}$
within $1\sigma$ (with the exception of 1.4$\sigma$ in the case of the
J $= 5 - 4$ line and with $1\sigma$ error bars in excess of 100 km
s$^{-1}$ for the J $= 7 - 6$ and $8 - 7$ lines).  We chose to fix the CVC line width
(and not the Broad component line width) because it is much more plausible that the disk
component has a line width independent of $J$ than the outflow.

We compared the reduced $\chi^2$ between single Gaussian and double
Gaussian fits and found clear evidence for outflow (Broad component)
in the CO $J = 5 - 4$ and $J = 6 - 5$ lines. The S/N in the cases of
the CO $J= 7 - 6$ and $J = 8 - 7$ was not high enough to
unambiguously distinguish the Broad and CVC components, i.e.,
the reduced $\chi^2$ was the same for single and double Gaussian
fits. We report the line widths and integrated line fluxes for the
Broad component and CVC in Table 3. 
For comparison, Alatalo et al. (2011) measured CVC line widths of
114 to 134 km s$^{-1}$ for $J = 1 - 0$ to $J = 3 - 2$ (with uncertainties up to
a few km s$^{-1}$) and adopted a single line width of $353 \pm 17$ km s$^{-1}$ for the Broad component from a fit to the $J = 2 - 1$ line.  Our CVC width is comparable, but our
Broad component linewidth differs by almost $4\sigma$; it is possible that the discrepancy
arises from different treatment of the baselines.
The lines fluxes were converted
from K to Jy km/s using the conversion factors given in the HIFI
Observers' Manual (herschel.esac.esa.int/Docs/HIFI/pdf/hifi\_om.pdf, p. 73, Table 5.5).
We assumed that the molecular emission region in NGC 1266 is compact
compared to {\it Herschel's} beams, which is supported by Alatalo et al. (2011).

\section{Dust Modeling}

The dust emission was modeled both to obtain the dust properties
(luminosity, temperature, and mass) and to derive the
wavelength-dependent dust optical depth for an extinction correction
to the high$-J$ CO lines. The dust parameters were fit using the
nested sampling algorithm PyMultiNest (Feroz and Hobson 2008; Feroz et al. 2009; Feroz et al. 2013; Buchner et al. 2014) with the method presented in Kamenetzky et al. (2014; hereafter K14); the
photometric data are in Table 4. We adopt the Casey (2012) model which
sums a grey-body distribution and a mid-infrared power-law
distribution with an exponential cutoff to account for warm dust. The
free parameters are $T$ (temperature, K), $\beta$ (emissivity index),
$\lambda_0$ (wavelength at which optical depth is unity, $\mu$m), and
$\alpha$ (power-law index). We assumed that calibration errors were
50\% correlated between measurements in different bands of the same
instruments (neither 0\% nor 100\% were appropriate and
experimentation showed that the parameters were not highly sensitive
to intermediate assumptions).

The dust-fitting results
and best-fit to the dust spectral energy distribution are given in
Table 5 and shown in Figure 3, respectively, and the likelihood distributions for the fitted parameters are in Figure 4. 
The dust is warm, with a small temperature uncertainty, $56
\pm 3$ K, although the uncertainties in $\lambda_0$ and $\beta$ are
larger $(\sigma_{\lambda_0} = 25~\mu$m and $\sigma_{\beta} = 0.29$) and
these are what most affect the CO extinction correction (see Section
4). The luminosity, Log $L_{\mathrm{dust}}/L_{\odot} = 10.44 \pm
0.01$, is well constrained. Using a dust opacity of
$\kappa_{\mathrm{125}\mu\mathrm{m}} = 2.64$ m$^2$ kg$^{-1}$ (Dunne et
al. 2003), the derived dust mass is Log $M_{\mathrm{dust}}/M_{\odot}
= 6.34 \pm 0.04$, corresponding to Log $M_{\mathrm{gas}}/M_{\odot} =
8.34 \pm 0.04$ for a gas-to-dust mass conversion factor of 100.
The quoted uncertainties account only for the statistical uncertainties in the derived gas mass.  The dust opacity is probably uncertain to a factor of approximately two, which propagates to an uncertainty of 0.3 dex in the dust mass.  Furthermore, the dust-to-gas mass conversion factor is probably uncertain up to a factor of two, yielding a net systematic uncertainty of $\sim0.5$ dex in the derived gas mass, dominating over the statistical uncertainty.

\section{Non-LTE CO Modeling and Gas Parameters}

To determine the gas physical conditions, we used a custom version of the RADEX non-LTE code
(van der Tak et al. 2007) combined with a Bayesian 
analysis (see Rangwala et al. 2011, Scott et al. 2011, Kamenetzky et
al. 2012, and Kamenetzky et al. 2014).  RADEX computes the intensities
of molecular lines by iteratively solving for statistical equilibrium
using an escape probability formalism. 
The inputs to RADEX are the gas density ($n_{H_2}$; H$_2$ is the main
collision partner for CO), the kinetic temperature ($T_{kin}$), and
the CO column density per unit line width ($N_{CO}/\Delta v$), which
sets the optical depth scale.  We also vary the area filling factor
($\Phi_A$).  The code generates a grid of model CO spectral line
energy distributions (in background-subtracted, Rayleigh-Jeans antenna
temperatures) for a large range of input parameters. From the marginalized best-fit parameters (again using PyMultiNest), we 
calculate the likelihoods of a set of secondary parameters, such as
pressure and molecular gas mass.

Generally two temperature components are required to fit the spectral line energy distributions; see the discussion in
K14. Motivated by the empirical result that the
ratio of mass of the warm gas component to the cool gas component
traced by CO in seventeen infrared-luminous nearby galaxies is $0.11
\pm 0.02$ (including NGC 1266, but the results are essentially
unchanged if NGC 1266 is excluded; K14), the mass
of the warm component was required to be less than the mass of the cool component.  Based on interferometry from Alatalo et al. (2011), a
source size of 22\arcsec~was assumed for the CVC and Broad emitting regions (this
constraint is revisited in \S 5), with areal filling factors $\le 1$, which allows the sources
to take on any sizes up to 22\arcsec~to find the model parameters that best represent
the data.
A
CO-to-H$_2$ abundance ratio of $3 \times 10^{-4}$ was assumed to scale
the CO column density to a total molecular gas column density (Lacy et al. 1994; the derived molecular gas masses scale inversely with the assumed value).  CO-to-H$_2$ abundance 
ratios of $1 \times 10^{-4}$ to $3 \times 10^{-4}$ are assumed in the literature. 
We have chosen the high end of this range, which is as large as the CO abundance can get for gas-phase abundances similar to the Solar neighborhood (most of the gas-phase carbon would be in the form of CO), because it makes our mass estimates conservative in the sense of producing
minimal molecular gas masses.

We did two types of RADEX model fitting to the CO spectral line energy
distributions: (1) using only the lines up to and including $J = 8 -
7$, in which the CVC and Broad components could be distinguished with
the spectral resolutions afforded by HIFI and ground-based heterodyne
observations, and (2) using the CVC model level populations
to generate line fluxes for the $J = 9 - 8$ and higher lines for
subtraction from the FTS fluxes to estimate and model the Broad
component fluxes. Case (2) is included because additional line flux
information is available in the FTS spectra: the CVC and Broad
components must sum to match the FTS line fluxes (within the
photometric uncertainties).  The CVC line fluxes, rather than the
Broad component fluxes, were subtracted because they were better
constrained for more of the low and mid-$J$ lines. To properly treat
the uncertainties, the likelihood distributions for the $J = 9 - 8$
and higher lines from the CVC model were used for subtraction from the FTS line fluxes,
thereby marginalizing over the ranges of allowed values.  
We explore Case (2) in the Discussion (\S 5) because it makes use of 
all available information and assumes there are no non-physical discontinuities
in the spectral line energy distributions of the different components.
Within the two broad
categories of models, various additional models were run to test for
the effects of extinction and assumptions about multiple components.

We first consider case (1), fitting only up to and including $J = 8
- 7$, for the CVC, with the constraint that the mass of the cold
component has to be greater than the mass of the warm component and with
no extinction correction applied (Fig. \ref{fig5} and Table 6). Not surprisingly, given what has
been observed for other galaxies, two temperature components are
required to simultaneously account for the lowest$-J$ lines ($1 - 0$
and $2 - 1$) and the mid$-J$ lines\footnote{Although only two
  components are required to fit the spectral line energy distribution
  well, it is likely that there are a continuum of temperatures and
  densities in the NGC1266 molecular interstellar medium. K14 has a discussion of the numbers of components.
  Further work exploring, e.g., models with power-law distributions of
  temperature, density, and column density is needed.  For the time
  being, we acknowledge that a continuum of physical conditions is
  likely present but also conclude that a cold component is required
  to explain the low$-J$ line fluxes and a higher-excitation component
  is required to explain the mid$-J$ lines -- a single component
  cannot simultaneously do both.}. The cold component is quite cold,
$6^{+11}_{-1}$ K (error bars are $\pm 1\sigma$ unless otherwise
indicated, where $\pm 1\sigma$ encompasses 68\% of the likelihood
distributions about the means), and the warm component is
$74^{+134}_{-26}$ K. The densities of the two components are not well constrained, both with appreciable likelihood from $10^2$ to $10^7$ cm$^{-3}$.  Because the density likelihoods are so broad, the pressures
(the products of the temperatures and densities) are poorly
constrained.\footnote{Temperature and density are typically degenerate
  over a fairly broad range in the collisional excitation of CO,
  although their product, the pressure, is {\it usually} better
  constrained because the temperature and density are anticorrelated in fits: 
  temperatures and densities can trade off to yield similar pressures -- low temperatures and high densities, and vice versa.} For the cold component the pressure likelihood is
appreciable from $10^3$ to $10^8$ K cm$^{-3}$. The pressure of the
warm component has a median of $10^{6.3}$ K cm$^{-3}$
and a $\pm 1\sigma$ range of approximately one dex.
The luminosities, which are largely independent of the model
parameters, are well separated and well constrained, with the cool
component (Log $L/L_{\odot} = 5.44^{+0.22}_{-0.18}$) required to explain the luminosities of the lowest few lines
and the warm (Log $L/L_{\odot} = 6.90^{+0.16}_{-0.16}$) component almost entirely responsible for the
luminosities of the lines above $J = 5 - 4$. Only about 3\% of the
luminosity emerges in the low$-J$ lines, which is on the low end of
the range observed in K14, but not extraordinary. As with density and
temperature, column density and beam filling factor are degenerate and
anticorrelated in fits; thus, we focus on their product, the beam-averaged
column density, which is converted to molecular gas mass as in K14. The cold and
warm component masses are Log $M/M_\odot = 8.3^{+0.5}_{-0.4}$ and Log
$M/M_\odot = 7.3^{+0.5}_{-0.5}$, yielding a cold-to-warm component
mass ratio of 0.1, similar to the average of the K14 sample ($0.11 \pm 0.02$).

With optically thick dust at far-infrared wavelengths
($\tau_{100\mu\mathrm{m}} = 4.7\pm 1.9$), the high$-J$ CO lines are
likely subject to extinction. Under the assumption of well-mixed
molecular gas and dust, the extinction correction can be applied by
dividing the line fluxes by a factor $(1 -
e^{-\tau_{\lambda}})/\tau_{\lambda}$. 
The correction has the effect of
slightly enhancing the bump in the spectral line energy distribution
starting at $J = 4 - 3$ and extending to $J = 8 - 7$.  Predictably,
the luminosity of the warm component rises slightly, from $7.9 \times 10^6$ to $9.7 \times
10^6~\mathrm{L_{\odot}}$, and the cold component luminosity is
virtually unchanged (where the luminosities include the lines from $J = 1 - 0$ to $8 - 7$).
Conversely, the pressure in the warm component drops by
about a factor of five and is more tightly constrained, indicating
that with the extinction correction the gas is better represented by a
single temperature component. Similarly, the cold component pressure
also drops and is likewise more tightly constrained.
The masses of the two components do not change significantly. Thus, the overall effect of the extinction
correction is to raise the CO (warm component) luminosity by
approximately 23\%, restrict the pressure likelihoods to smaller
ranges (despite the additional uncertainties introduced by the
uncertainties in the dust optical depths at the line frequencies), and
leave the masses unchanged.  For most of the remainder of the analysis, we 
consider the non-extinction corrected results because the properties of
dust in the outflow are unknown and may not be similar to the 
general dense interstellar medium; however, we do describe the implications of
making the extinction correction where they are relevant.

CARMA CO $J= 1 - 0$ interferometry resolves the CVC into compact
emission (``nucleus''), with a deconvolved radius of 60 pc, and an
extended region (``envelope'') at least twice that size (Alatalo et al
2011). We are unable to obtain fits of the CVC spectral line energy distribution that are
consistent with the $J = 1 - 0$, $2 - 1$, and $3 - 2$ lines if we
restrict the emitting region to a radius of 60 pc: the line fluxes are
underproduced by the models and a substantially larger radius must be
allowed, confirming the presence of the envelope. To attempt to quantify
the fraction of low$-J$ emission arising in the nucleus and envelope,
we ran models with the same percentages of $J = 1 - 0$, $2 - 1$, and
$3 - 2$ lines subtracted from the CVC fluxes and a compact source
radius fixed at 60 pc, such that the fractions of the line fluxes
subtracted represent the flux arising from the envelope. The model
fits are not unique, but we conclude that likely at least 50\% of the
CVC flux from the $J= 1 - 0$ and $J = 2 - 1$ lines arises from the
envelope.\footnote{It is unlikely that missing flux in the CARMA CO $J = 1 - 0$ could substantially affect this comparison because Alatalo et al. (2011) measured very similar $J = 1 - 0$ line fluxes with CARMA and the IRAM 30 m.}

Only a single gas component is required to fit the Broad component
spectral line energy distribution up to and including the $J = 8 - 7$ line (Fig. \ref{fig6} and Table \ref{Broad_all}): while the temperature is low 
(median value of 34 K) and
well constrained ($1\sigma$ range of 30 K to 41), the density
likelihood distribution is very broad (partially because of the large uncertainties
in the mid$-J$ Broad component line fluxes), 
ranging from $10^4$ to just over $10^7$ cm$^{-3}$, leading to a broad
pressure likelihood distribution ($3 \times 10^5$ K cm$^{-3}$ to $1
\times 10^9$ K cm$^{-3}$). In fact, the density and pressure have uniform likelihood to the high (unphysical) end of the model grid, indicating unconstrained upper limits.  Hence, while it seems likely that the gas
is cold, its pressure and excitation are uncertain owing to the large
allowed range in density.  
The mass of the Broad component is well constrained with $M_{\mathrm
Broad} \sim 4 \times 10^7~\mathrm{M_{\odot}}$, intermediate between the cold
and warm CVC masses.
Thus, the mass and luminosity ($6.9 \times 10^6~\mathrm{L_{\odot}}$) of the
Broad component are well constrained, but the gas pressure is not. 
 As described next, using the full span of lines, including the higher$-J$, lines indicates both cold and warm components, with the cold component parameters being very similar to those derived using only the low and mid$-J$ lines.

In case (2) including the high-$J$ lines, the CVC $J= 9 - 8$ to $J = 13 - 12$ model line fluxes are
subtracted from the FTS line fluxes to derive the high$-J$ Broad
component line fluxes. The Broad component now requires two
temperature components to produce physical fits to the spectral line energy distribution (Fig. \ref{fig7} and Table \ref{Broad_all}).
\footnote{A single component yields solutions at low or high
temperatures, depending on what lines are included; removal of the
mass constraint $M_{\mathrm warm} < M_{\mathrm cold}$ yields an
insignificant warm component and extremely broad likelihood
distributions for the cold component, which cannot alone account for a
sufficiently broad range of physical conditions to fit the spectral line energy distribution.}
The spectral line energy distribution is shown in Fig. \ref{fig8}.  With two components
and the mass constraint, 
the luminosities of the cold and warm
components are $L_c \sim 3.0 \times 10^6~\mathrm{L_{\odot}}$
and $L_w \sim 1.7 \times 10^7~\mathrm{L_{\odot}}$, and the temperatures are well separated, with maximum likelihood values of $\sim30$ K and $\sim$ 150 K, and masses of approximately Log $M/M_{\odot} = 7.6$ and $6.3$, respectively.
Thus, the cold component dominates the mass and the warm
component dominates the luminosity. 
The warm H$_2$ density distribution is bimodal, with appreciable likelihood in the $10^{3.5} - 10^5$ cm$^{-3}$ range and above $10^6$ cm$^{-3}$, but low likelihood between those ranges.  The high range runs to the grid limit, but we disregard these densities ($n > 10^6$ cm$^{-3}$) because they are unphysically high.  The high pressure ($nT > 10^8$ K cm$^{-3}$) are also unphysical because they correspond to the unphysical densities.
While the extrapolation of the CVC to high-J introduces substantial
uncertainty, the Broad component solutions are qualitatively similar
to what has been observed for most other galaxies that have been
studied in similar detail: two temperature components are required,
the warm component is responsible for the bulk of the CO luminosity,
and most of the mass is in the cold component. We reiterate that the
warm Broad component parameters are meaningful only if our
extrapolation of the separation into Broad and CVC components to the
$J=9-8$ and higher lines, based on the case (1) model, is
valid. 
However, the temperature of the cooler component, $\sim$ 60 K, is consistent with the analysis of the $J = 8 -7$ and below lines only. 
Applying the extinction correction drives up the $J= 8 - 7$ and
higher line fluxes and hence the warm Broad component luminosity to $3.5\times 10^7~L_\mathrm{\odot}$, an increase of approximately 2 (where the luminosity now includes all of the lines from $J = 1 - 0$ to $J = 13 - 12$). 

Our rotational CO molecular gas mass measurements can be compared to mass measurements from H$_2$ observations.  Roussel et al. (2007) used vibrational transitions of H$_2$ to measure the molecular gas mass in the nuclear region of NGC1266 with an irregular 287 square arcsecond aperture.  They measured gas masses of $1.3 \times 10^7~\mathrm{M_{\odot}}$ and $2.2 \times 10^6~\mathrm{M_{\odot}}$ in the temperature ranges $200~\mathrm{K} < T < 400~\mathrm{K}$ and $400~\mathrm{K} < T < 1,450~\mathrm{K}$, respectively. They did not have sufficient velocity resolution to separate the CVC and Broad components, both of which presumably contribute to their line fluxes.  Using vibrational H$_2$, they did not probe the gas in our cool component (probed with rotational transitions of CO).  Their $200~\mathrm{K} < T < 400~\mathrm{K}$ component overlaps substantially with our warm Broad component ($T = 146^{+974}_{-27}$ K) and somewhat with our warm CVC component ($T = 70^{+130}_{-26}$ K, where the error bars are $1\sigma$), which yield a combined mass of $2.2 \times 10^7~\mathrm{M_{\odot}}$, approximately $1.7\times$ the H$_2$ mass.  These mass estimates compare favorably considering that much of our CVC warm component could be cooler than 200 K.  (The $T > 400$ K H$_2$ component mass of $2.2 \times 10^6~\mathrm{M_{\odot}}$ may also overlap with our Broad component, which has non-zero temperature likelihood  up to $\sim1,000$ K.)  Thus, the H$_2$ and CO-derived masses provide consistency checks, which the temperature estimates pass, but do not provide direct comparisons.

\section{Discussion}

NGC 1266 is notably complex for an S$0$ galaxy: there are at least
five discernible components of CO emission from the nuclear region:
(1) and (2) the CVC, presumably a disk, comprised of cool and warm components and with mass
dominated by the cool component, but in which the warm component
dominates the luminosity; (3) an extended, cool CVC envelope of radius
$r \gtrsim 100$ pc; and (4) and (5) the cool (more massive) and warm
(more luminous) Broad components associated with the outflow. In fact, more components (or a
continuum of physical conditions) are likely present, but they cannot
be distinguished with the data in hand. Here we compare estimates of
the physical conditions of the various components, with a focus on
masses, and discuss the gas excitation mechanisms.

\subsection{Physical Conditions of the CO Emission Components with a 
Focus on Masses}

Our modeling indicates cool and warm molecular gas masses for the CVC
-- including the envelope -- of $M_{\mathrm CVC,c} \simeq 2\times 10^8
~\mathrm{M_\odot}$ and $M_{\mathrm CVC,w} = 2\times 10^7~\mathrm{M_\odot}$, so the CVC
mass is dominated by the cold ($\sim 10$ K) component. For comparison,
Alatalo et al. (2011) used a CO conversion factor $\mathrm{X_{CO}} = 2
\times 10^{20}~\mathrm{cm^{-2}}~\mathrm{(K~km~s^{-1})^{-1}}$ and peak
CO (J $= 1 - 0$) flux of 24.3 Jy km s$^{-1}$ beam$^{-1}$ to estimate
the molecular gas mass. They found a nuclear (i.e., excluding the
envelope) CVC mass of $M_{\mathrm CVC, nuc} \approx 4\times 10^8
~\mathrm{M_\odot}$ (the $\mathrm{X_{CO}}$ factor should be relatively
insensitive to the mass in the warm component). They use two
additional methods to determine the CVC mass. Firstly, along the major
axis in the position-velocity diagram they trace rotation out to a radius of $\sim
54$ pc and derive a maximum projected rotation velocity of 110 km
s$^{-1}$ ($34^{\circ}$ inclination), from which they measure an
enclosed dynamical nuclear mass of $M_{\mathrm dyn,nuc} \approx 5\times
10^8 ~\mathrm{M_\odot}$. Secondly, from an IRAM 30 m $^{13}$CO ($J = 1 - 0$)
line flux and a relation between $^{13}$CO line flux and extinction,
they derive $M_{\mathrm CVC, nuc}\approx 3 \times 10^8~\mathrm{M_\odot}$. By
assuming that the rest of the CVC mass not in the nucleus is in the
envelope, they estimate a mass of $M_{\mathrm{CVC,env}}
\approx 6\times 10^8~\mathrm{M_\odot}$ in the envelope. Their combined nucleus
and envelope mass of $M_{\mathrm CVC} \approx 10^9~\mathrm{M_\odot}$, is just
outside the $1\sigma$ upper bound of our model; hence it cannot be
considered inconsistent with our results. In fact, given the
uncertainties, these values are all remarkably consistent. Our new
observations and modeling have added to our understanding by showing
that 10\% of the CVC mass is in a warm, high pressure component ($nT =
10^{6.3}$ K cm$^{-3}$ with a $\pm 1\sigma$ range of approximately one
dex).

Estimation of the mass and physical conditions of the Broad component,
the molecular outflow from the nucleus of NGC 1266, is important
because it will yield insight into the acceleration of the outflow and
because the gas could enrich the halo of NGC 1266.  To
summarize the physical conditions we derived for the Broad component:
The low and mid$-J$ spectral line energy distribution is well fit by a
single component ($T = 30 - 40$ K, $M \sim 4 \times 10^7~\mathrm{M_\odot}$, $L
\approx 6.9 \times 10^6~\mathrm{L_\odot}$).  
Similarly, when the model CVC high-$J$ line fluxes are
subtracted from the FTS line fluxes to yield Broad component high-$J$ line fluxes, both warm ($L_{\mathrm broad,
w}\sim 2 \times 10^7~\mathrm{L_\odot}$, $P_{\mathrm broad, w} \sim 4 \times
10^7 ~\mathrm{K~cm}^{-3}$, $M_{\mathrm broad, w}\sim 2 \times 10^6~\mathrm{M_\odot}$) and cool ($L_{\mathrm broad, c}\sim 3 \times 10^6~\mathrm{L_\odot}$,
$M_{\mathrm broad, c} \sim 2 \times10^7~\mathrm{M_\odot}$ to $8 \times 10^7~\mathrm{M_\odot}$) Broad components are also required\footnote{The luminosity is 2 times larger with
the extinction correction applied.}. The low temperature
component is subdominant in luminosity.  
Alatalo et al. (2011) also used RADEX modeling for the three
lowest$-J$ CO lines, which showed (very non-uniquely) that the $J = 1
- 0$ to $3 - 2$ line fluxes were consistent with $n \sim
10^3~\mathrm{cm^{-3}}$, $T \sim 100$ K and $N\mathrm{(CO)} \sim
10^{16}~\mathrm{cm}^{-2}$, and an outflow mass of $2.4 \times 10^7~\mathrm{M_\odot}$ for a CO/H$_2$ abundance of $1 \times 10^{-4}$. Using our assumed relative CO abundance of $3 \times 10^{-4}$ instead, this would yield a molecular gas mass of $8 \times 10^{6}~\mathrm{M_\odot}$, which is not 
consistent with our result.  Our mass estimate 
includes more lines and determines the parameter likelihoods
rather than just finding {\it a} solution that fits the data. It is likely,
therefore, that the mass of the outflow is approximately one order of
magnitude smaller than the mass of the CVC.

In a recent paper, Alatalo et al. (2015) argue that the Broad component mass
is $2\times 10^8\, M_\odot$, nearly an order of magnitude larger than
their previous estimate, with important implications for the driving
of the outflow. This revised mass is based on the use of a CO
conversion factor commonly used for ULIRGs (Bolatto, Wolfire, \& Leroy
2013). Alatalo et al. adopt this conversion factor on the grounds that
the outflow must be optically thick in CO and dominated by dense gas,
as they detect high-velocity components in CARMA observations of the
HCN $J=1-0$ (blueshifted and redshifted) and CS $J=2-1$ (blueshifted
only) lines. This also raises their outflow mass estimate to 110
$M_\odot$ yr$^{-1}$. However, we are skeptical of this conclusion.
Even if the broad line emission is both optically thick and arises in
dense gas, that does not imply that the ULIRG conversion factor is
appropriate for determining the gas mass from the CO line flux.  Also, their revised outflow mass is comparable to our
estimate for that of the CVC component, which would mean that the
latter does not dominate the gas mass.

The dust mass (Table 5) also provides an estimate of the total mass of
the molecular gas. Assuming a gas-to-dust mass ratio of 100 (K14
argue it could be up to a factor of two smaller), our dust mass
converts to a gas mass of $2.2 \times 10^8~\mathrm{M_\odot}$. Combined, the
ensemble of measurements favors a total gas mass nearer to $10^8$ than
$10^9~\mathrm{M_{\odot}}$.  We note that the primary uncertainty in our molecular gas estimates is the relative abundance of CO to H$_2$, which likely contributes an uncertainty of approximately a factor of two to our molecular gas masses in addition to the statistical uncertainties in the tables and figures, and that the molecular gas masses may be biased low if anything because of our relative abundance choice ($3 \times10^{-4}$), but we are not subject to uncertainties in the CO X$_{\mathrm{CO}}$ factor or lack of knowledge of the CO rotational level excitation.

Alatalo et al. (2011) estimate a gas outflow rate of 13 $\mathrm{M_\odot}$ yr$^{-1}$, corresponding to a depletion timescale of $\tau_{\mathrm out} <
85$ Myr. The precision of our mass measurements significantly improve upon
this estimate and suggest that the gas depletion time may be even shorter. Using the conversion from total infrared luminosity (8
to 1000 $\mu$m) to star formation rate of Bell (2003),
if the far-infrared luminosity of NGC 1266 is due entirely to star
formation its star formation rate is approximately 6 $\mathrm{M_\odot}$ yr$^{-1}$. This corresponds to a gas depletion timescale ($\tau_{\mathrm sf}$) approximately
twice as long as $\tau_{\mathrm out}$, suggesting that much of the
reservoir will likely be ejected from the nuclear region before it can
be converted to stars.

\subsection{Gas Excitation Mechanisms}

The ratio of total CO to far-infrared luminosities, $L_{\rm CO}/
L_{\rm FIR}$, is anomalously large, $1.7\times 10^{-3}$, compared to
the norm of $4\times
10^{-4}$ found in the sample of K14.\footnote{$L_{\rm CO}/
L_{\rm FIR} =1.7\times 10^{-3}$ is extinction corrected, for comparison to the K14 result which is also extinction corrected.   For NGC 1266, $L_{\rm CO}/
L_{\rm FIR} =1.0\times 10^{-3}$ without an extinction correction.} In that sample it is exceeded only by that of NGC 6240, a near-ULIRG merger
with two nuclei, a buried AGN, and a strong case for shock-excited
warm molecular gas (Meijerink et al. 2013). M82 and Arp 220, both with luminosities dominated by
star formation, have CO spectral line energy distributions that peak around the $J = 7 - 6$ line,
then turn down for higher $J_{\rm upper}$ (Fig. \ref{fig9}). In contrast,
Mrk 231, a Seyfert 1 galaxy, has a CO spectral line energy
distribution (characteristic of AGN dominated galaxies) that remains
flat above $J = 7 - 6$ for several lines before declining more slowly
than starburst galaxies. NGC 1266's CO spectral line energy distribution is intermediate,
suggesting perhaps that either both star formation and the AGN
contribute to the CO excitation, or that the AGN is simply too weak to
dominate the excitation of the high$-J$ lines.

Using the additional HIFI observations reported here, we are able to
separate the CO luminosity of the Broad component from that of the CVC.  For
the CVC, the derived $L_{\rm CO}/L_{\rm FIR} =
3\times 10^{-4}$ is indistinguishable from the mean value in the
K14 sample. This implies that this ratio must be
substantially larger for the outflow, and possibly very large if the
far-infrared emission is dominated by the CVC. This has important
implications for the energetics of the outflow, as we discuss below.

There are multiple mechanisms that
could contribute to the heating and excitation of the warm, Broad (outflow) component of the gas:
photodissociation regions (PDRs) associated with high-mass star
formation regions; shocks from outflows, infall, or other gas
dynamics; cosmic rays; and an X-ray dissociation region (XDR) produced
by the AGN. Given the modest star formation rate and sub-dominance of
cosmic rays in the warm molecular gas heating of starburst galaxies
(e.g., M 82: Panuzzo et al. 2010; Arp 220: Rangwala et al. 2011),
cosmic rays are not a likely candidate (cosmic rays are still likely
an important source of heating of the cold component of the CVC and
even of the cold Broad [outflow] component, given the probable high optical
depth of the clouds). Pellegrini et al. (2013) considered three
excitation mechanisms for CO in NGC 1266: PDRs, XDRs, and shocks.
With only mid and high$-J$ line fluxes from {\it Herschel} SPIRE FTS
observations, which could not spectrally resolve the lines, they did not consider
the CVC and Broad components separately.
While a two-component PDR model ($n_{\mathrm{H}} =
10^{3.75}~\mathrm{cm^{-3}}$ and G$_0$ = $10^6$) and ($n_\mathrm{H} =
10^{5.5}~\mathrm{cm^{-3}}$ and G$_0 = 10^{3.5}$) could explain both
the CO spectral line energy distribution and the rotation lines of H$_2$ observed by {\it Spitzer},
Pellegrini et al. (2013) ruled this model out because it predicted a 
$L_{\rm CO}/L_{\rm FIR}$ ratio three orders of magnitude smaller than
the observed ratio.
XDRs were also ruled out based on the $L_{\rm CO}/L_{\rm FIR}$
ratio.\footnote{However, they misinterpreted Rangwala et al. (2011) to
argue against XDRs. Quoting from their paper: ``We do detect
absorption from H$_2$O$^+$, but this has been seen in systems where
XDR are ruled out (e.g., Arp220; Rangwala et al. 2011 ).''  In fact,
an XDR {\it is} present in Arp 220 --- it is required to explain the
high column density of molecular ions; however, the AGN is not the
primary heating mechanism for the molecular gas that is traced by
the CO emission.} Pellegrini et al. (2013) argued that the CO SLED was a
composite, with the lowest$-J$ lines produced in a PDR and the
higher$-J$ lines produced by C-shocks, which can also produce the
observed water emission. Meijerink et al. (2013) reached similar
conclusions about NGC 6240 from their analysis of the {\it Herschel}
SPIRE FTS CO spectral line energy distribution.

Here we revisit the possibility of PDRs as an important agent in the
heating of the warm molecular gas, using the additional information
obtained by resolving the CVC and Broad components in the $J= 1 - 0$
to $J = 8 - 7$ lines with ground-based and HIFI
observations. Specifically, we consider the $L_{\rm CO}(J = 6 -
5)/L_{\rm FIR}$, $L_{\rm CO}(J = 6 - 5)/L_{\rm CO}(J = 1 - 0)$,
$L_{\rm CO}(J = 11 - 10)/L_{\rm CO}(J = 6 - 5)$, and $L_{\rm CO}(J =
11 - 10)/L_{\rm CO}(J = 1 - 0)$ ratios, combined with the number
density constraints from our non-LTE modeling. The PDR models are
those of Wolfire et al. (2010) (M.~Wolfire, private communication 2011).

The results are shown in Figure \ref{fig10}. Consider first the $L_{\rm
CO}(J = 6 - 5)/L_{\rm FIR}$ ratio (Figure \ref{fig10} upper left). The green and
blue curves show the observed values for the CVC and Broad components,
respectively, while the dashed vertical lines show the allowed density
regimes, similarly color-coded. The CVC and
Broad line-to-continuum ratios are nearly identical, reflecting the
similarity of the fluxes in the two components. For both the warm CVC
and Broad components, we find that PDR solutions {\it are} acceptable
for $G_0$ between $10^3$ and $10^4$. However, there is an important
caveat to this conclusion. In the absence of high-resolution spectral
and continuum imaging, it is not known what fraction of the
far-infrared luminosity should be assigned to each component. The
observed ratios have therefore been calculated by assigning 100\% of
the far-infrared flux to both. It is highly likely that this results
in a significant underestimate of the true line-to-continuum ratio for
the Broad component. As discussed above, the total $L_{\rm CO}/L_{\rm
FIR}$ ratio for the CVC component is typical for star-forming galaxies
(K14) when all of the far-infrared luminosity is
assigned to this component, while for the Broad component this ratio
is unusually high. Combined with the other evidence for star formation
in this galaxy (Alatalo et al. 2011; Alatalo et al. 2014), this suggests
that a large fraction of the far-IR arises from the CVC component. If
only a fraction $f_B$ should be associated with the Broad component emission,
then the acceptable Log $G_0$ range in Figure \ref{fig10}a will shift down
by approximately $\log(1/f_B)$ (for $f_B \gtrsim 3\%$), implying that
$G_0 < 10^3$ for $f_B < 10\%$.  This has important implications when we
consider the other PDR diagnostics.

Figure \ref{fig10} upper right compares the $L_{\rm CO}(J = 6 - 5)/L_{\rm CO}(J = 1
- 0)$ ratios for the two components to the PDR model predictions. We
still find acceptable solutions for $G_0$ in the $10^3 - 10^4$ range;
for the CVC, the PDR models are pushed to the low end of the density
range, while for the Broad component acceptable solutions for both
diagnostics require $G_0\sim 10^4$ and densities at the upper end of
the range allowed by the likelihood analysis. 

The line ratios plotted in the remaining two figure panels must be regarded
as uncertain, since the division of the $J = 11 - 10$ line flux into
CVC and Broad components relies on the extrapolation of the CVC model
spectrum to lines for which we have only unresolved FTS fluxes. 
However, the constraints imposed by the $L_{\rm CO}(J = 11 - 10)/L_{\rm CO}(J =
6 - 5)$, and $L_{\rm CO}(J = 11 - 10)/L_{\rm CO}(J = 1 - 0)$ ratios
are similar, for both components. For the CVC emission, the solutions
imposed by matching the $J = 11 - 10$ line ratios are somewhat at odds
with the other PDR constraints: although a similar range of $G_0$ is
allowed, satisfying the high$-J$ line ratios require densities that
are much larger (an order of magnitude or more) than for the
line-to-continuum and $L_{\rm CO}(J = 6 - 5)/L_{\rm CO}(J = 1 - 0)$
ratios. The CVC $J = 11-10$ line flux would have to be much smaller
than our extrapolated value to vitiate this conclusion. For the Broad
component, the density constraints can be simultaneously satisfied for
all the observed ratios, but the high$-J$ line ratios require $G_0
\gtrsim 10^4$. If $f_B \ll 1$, as seems likely, then there would be no
PDR model parameter space in which all of the constraints can be
met. As with the CVC component, there will be acceptable solutions if
the Broad $J = 11 - 10$ line flux is much smaller than our estimate,
but if this is true, then this rules out a PDR origin for the CVC
component, as it must produce the remainder of the line flux. Thus,
independent of our modeling, at least one of the two components cannot
be explained as PDR emission.

In summary, the mid$-J$ and low$-J$ lines from both the CVC and Broad
components can in principle arise in PDRs, provided $G_0 \lesssim
10^4$. However, the higher$-J$ emission, as represented by the $J =
11-10$ line, does not fit easily into this picture, and at least one of the
two components must contain line contributions from a non-PDR source,
independent of our decomposition of the $J = 11-10$ line flux. We
therefore conclude that shocks are likely a substantial and possibly
dominant heating mechanism for the warm molecular gas in the CVC and
Broad components. This seems particularly likely for the Broad
component, given the identification of this component with the
molecular outflow.

X-ray dissociation regions (XDRs) have been found to be the likely excitation source of warm nuclear CO
emission in galaxies with AGN (e.g., Meijerink et al. 2013). Is it
possible
that the AGN in NGC1266 powers the CO emission via an XDR?  A definitive
analysis of CO emission from an XDR is not possible because the molecular
gas densities and temperatures are not constrained tightly enough and
because the AGN X-ray luminosity is uncertain.  While it is possible that
the $L_\mathrm{CO}/L_\mathrm{FIR}$ ($L_\mathrm{CO}/L_\mathrm{Bolometric} = 1 \times 10^{-5}$ and $2.9 \times 10^{-4}$ for the CVC cold and warm
components, respectively, and $L_\mathrm{CO}/L_\mathrm{Bolometric} = 1.1 \times 10^{-4}$ and $6.3 \times 10^{-4}$ for the Broad cold and warm
components, respectively, not extinction corrected) could be accommodated by an XDR (see Bradford et al. 2009), a
substantial fraction of CO emission arising from an XDR seems 
unlikely
because (a) the AGN would have to dominate the bolometric luminosity of the galaxy,
for which there is presently no observational support, and (b)
 the conversion of hard X-ray luminosity to CO
luminosity would have to be almost maximally efficient ($\sim$1\%).

\section{Conclusions}

New {\it Herschel Space Observatory} HIFI data enable us to spectroscopically resolve the nuclear central velocity component (CVC; 123 km s$^{-1}$ FWHM) and Broad (outflow; 278 km s$^{-1}$ FWHM) components of CO $J = 5 - 4$, $J = 6 - 5$, $J = 7 - 6$, and $J = 8 - 7$ lines for the first time.  Modeling of the $J = 1 - 0$ to $J = 8 - 7$ CVC reveals two components:  cold ($T \sim 10$ K), comprising most of the mass ($2 \times 10^8~\mathrm{M_{\odot}}$) and a minority of the luminosity ($3$\%), and warm ($T \sim 100$ K), with a minority of the mass ($2 \times 10^7~\mathrm{M_{\odot}}$) and a majority of the luminosity (97\%).  The CVC $L_\mathrm{CO}/L_\mathrm{FIR}$ is typical of star-forming galaxies and consistent with a modest star formation rate of 6 $\mathrm{M_{\odot}}~\mathrm{yr^{-1}}$.  With an estimated outflow rate of 13 $\mathrm{M_{\odot}}~\mathrm{yr^{-1}}$ the reservoir would be depleted before it could be converted to stars at the current consumption rates, but assuming that the CVC feeds the outflow perhaps 1/3 of the mass is destined to form stars.  Our models fail to fit the CVC CO spectral line energy distribution in the lowest few lines if the radius is $< 60$ pc, hence we conclude that on order 50\% of the emission arises from a larger region, consistent with the previous conclusions of Alatalo et al. (2011).

For the Broad component identified as an outflow, as for the CVC, the gas densities and pressures were not well constrained, partially because of the relatively large number of temperature components and uncertain velocity decomposition for the high-J lines.    Subtraction of the CVC J = 9 - 8 and above model line fluxes from {\it Herschel} SPIRE FTS line fluxes to estimate the Broad component line fluxes yields strong evidence that it also has warm and cold components, with again nearly an order of magnitude more CO luminosity arising from the warm component and masses of $4 \times 10^7~M_{\odot}$ (cold) and $2 \times 10^6~M_{\odot}$ (warm).  Unlike the CVC, the outflow $L_\mathrm{CO}/L_\mathrm{FIR}$ ratio is anomalously high.  Extinction corrections under the assumption of well-mixed dust and molecular gas raise the outflow luminosity as much as a factor of 2 but do not have a strong effect on the derived masses.

Comparison of the CO spectral line energy distribution, far-infrared luminosity, and CO modeling results to photodissociation models indicate that the warm components of both the CVC and outflow cannot be excited by PDRs; while just one of them could be, it is likely that both warm components are shock excited (especially the outflow).  Thus, our analysis provides strong evidence for the warm outflow gas (which is responsible for the bulk of the CO luminosity and anomalously high $L_\mathrm{CO}/L_\mathrm{FIR}$ ratio) being shock excited and driven by the AGN since the star formation is too anemic.

NGC 1266 remains enigmatic because the nuclear gas, with an apparently short lifetime because of the outflow and star formation, does not have a clear origin.  High angular resolution dust continuum interferometry will be required to further clarify the molecular gas physical conditions and various components, although large-scale mapping (several 100 pc) may be required to further elucidate the origin of the molecular gas.

\acknowledgments

We thank the anonymous referee for a thoughtful and constructive report.  We also thank Alex Conley for helpful discussions of modeling.  SPIRE has been developed by a consortium of institutes led by Cardiff
Univ. (UK) and including Univ. Lethbridge (Canada); NAOC (China); CEA,
LAM (France); IFSI, Univ. Padua (Italy); IAC (Spain); Stockholm
Observatory (Sweden); Imperial College London, RAL, UCL-MSSL, UKATC,
Univ. Sussex (UK); Caltech, JPL, NHSC, Univ. Colorado (USA). This
development has been supported by national funding agencies: CSA
(Canada); NAOC (China); CEA, CNES, CNRS (France); ASI (Italy); MCINN
(Spain); SNSB (Sweden); STFC, UKSA (UK); and NASA (USA).This research
has made use of the NASA/IPAC Extragalactic Database (NED) which is
operated by the Jet Propulsion Laboratory, California Institute of
Technology, under contract with the National Aeronautics and Space
Administration.  This work is supported in part by NASA award 1432844.  JRK gratefully acknowledges the support of an NSF Graduate Fellowship.

{\it Facilities:} \facility{Herschel Space
Observatory}

\clearpage

\begin{figure}
\epsscale{1.0}
\plotone{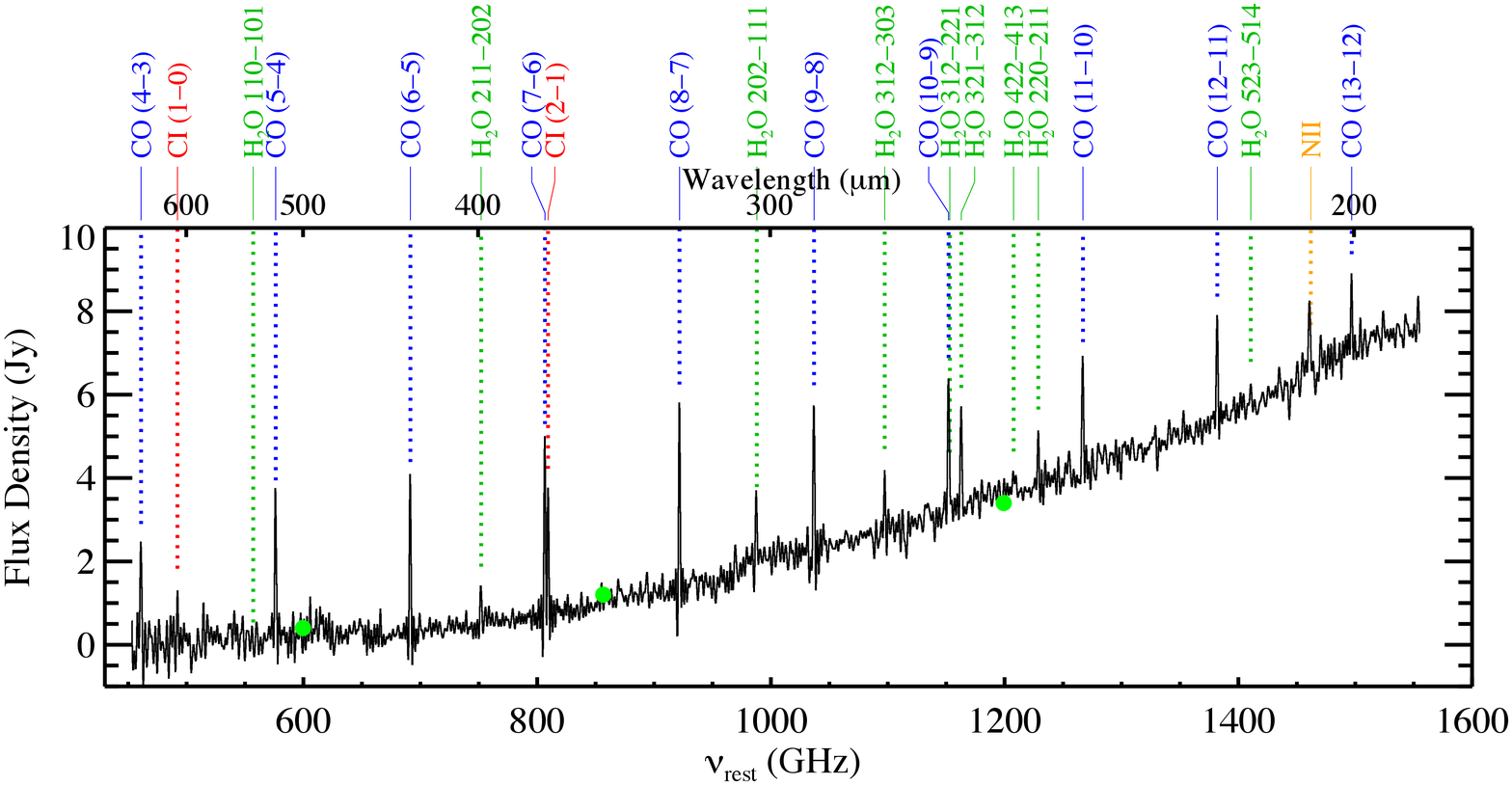}
\caption{Combined {\it Herschel} SPIRE FTS Spectrometer Short-Wave and Spectrometer Long-Wave spectra.
  The dust continuum is apparent rising to higher frequencies and prominent emission lines are labeled.  Filled green circles at approximately 600 GHz, 860 GHz, and 1220 GHz indicate the photometry of SPIRE PLW, PMW, and PSW, which are consistent with the spectroscopic flux densities.\label{fig1}} 
\end{figure}

\clearpage 

\begin{figure}
\epsscale{1.0}
\plotone{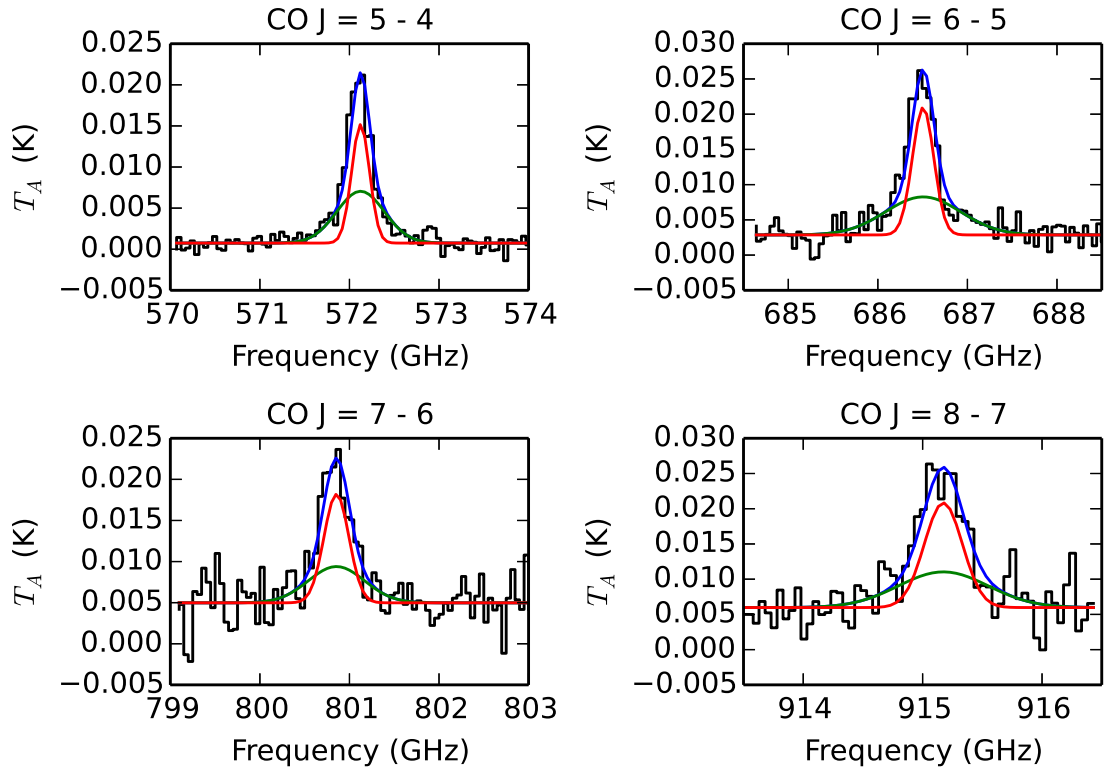}
\caption{{\it Herschel} HIFI CO spectra and their decompositions into CVC (red) and Broad (outflow, green) components.  The blue lines are their sums.  \label{fig2}}
\end{figure}

\clearpage 

\begin{figure}
\epsscale{1.0}
\plotone{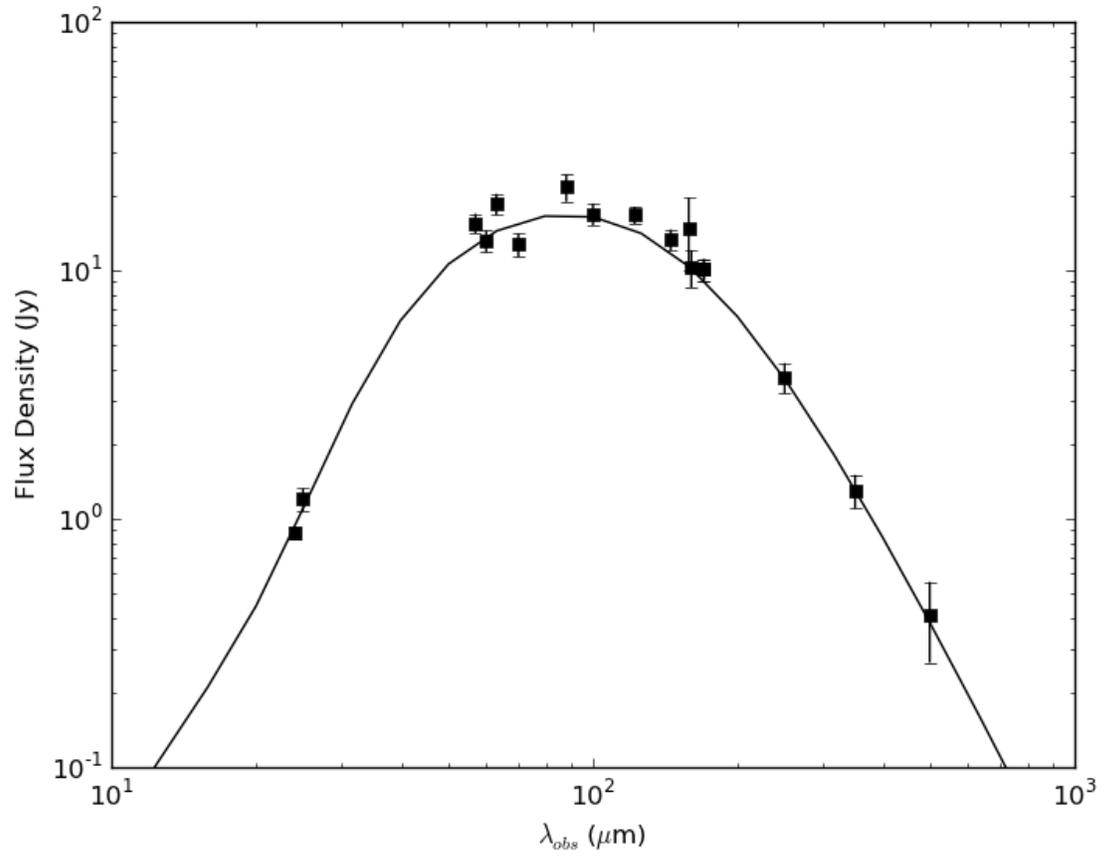}
\caption{Continuum photometry (listed in Table 4) and best-fit dust spectrum.  \label{fig3}} 
\end{figure}

\begin{figure}
\epsscale{1.0}
\plotone{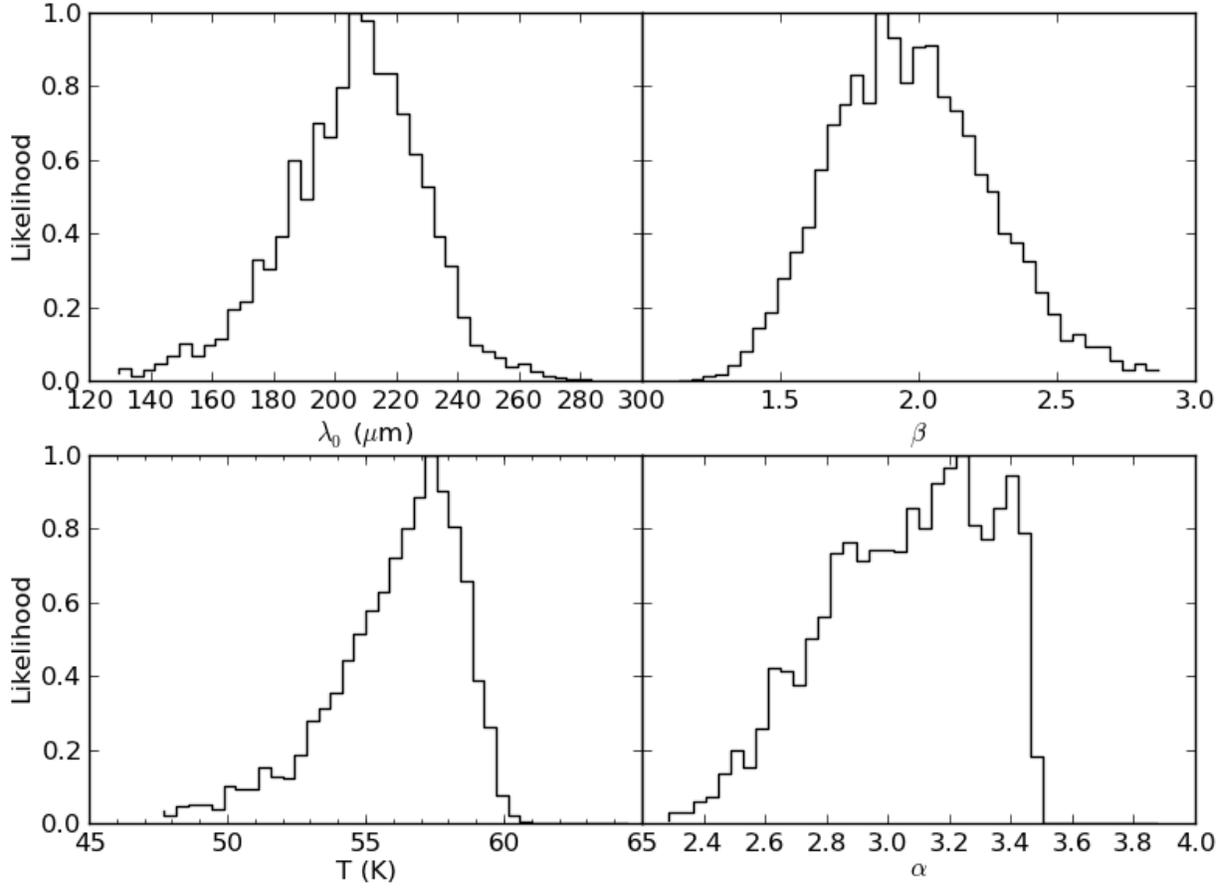}
\caption{Best-fit dust parameter likelihood distributions:  $\lambda_0$, wavelength at which the optical depth is unity, $\beta$, spectral index of emissivity, temperature $T$, and Wien-side spectral index $\alpha$.  \label{fig4}} 
\end{figure}

\begin{figure}
\plotone{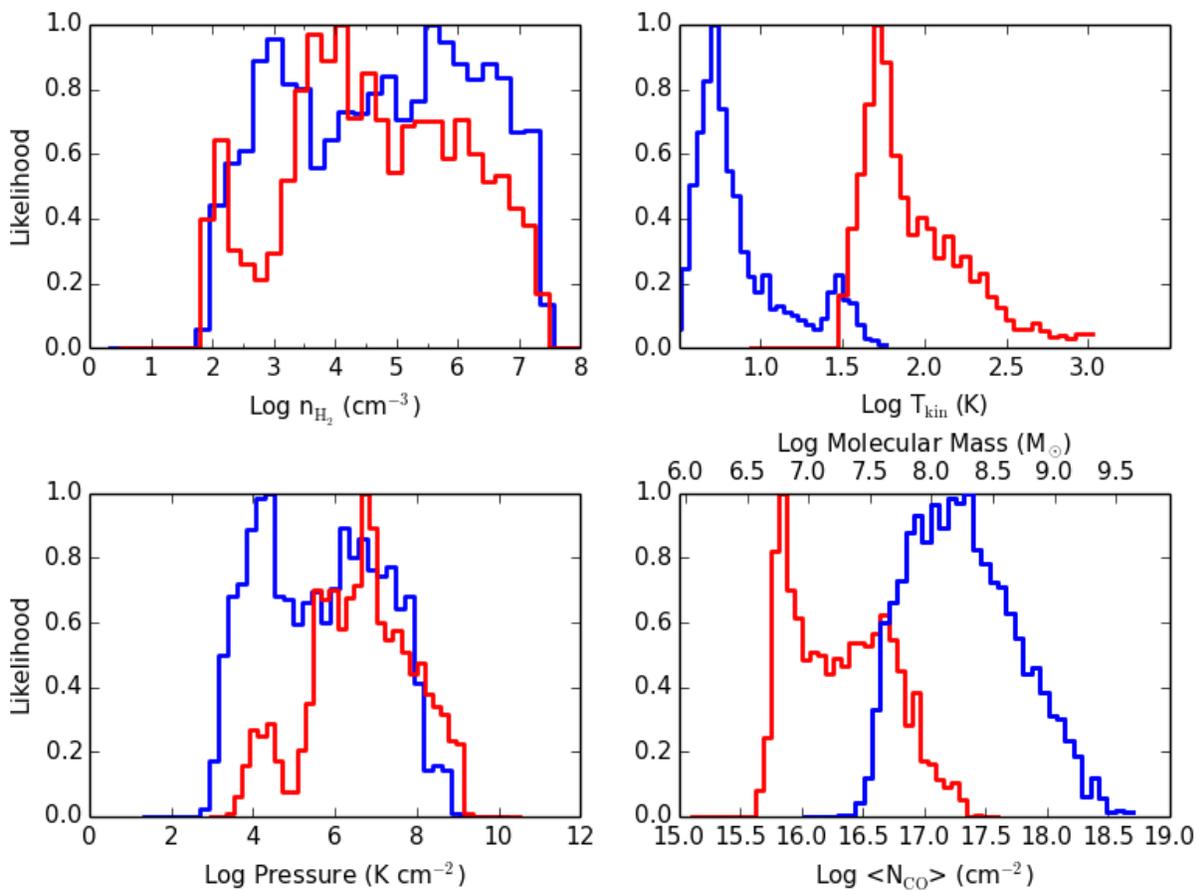}
\caption{Marginalized parameter likelihoods for the CVC cold (blue) and warm (red) components, subject to the constraint that $\mathrm{M_{warm} < M_{cold}}$ and using CO lines from $J = 1 - 0$ to $J = 8 - 7$. \label{fig5}} 
\end{figure}

\begin{figure}
\plotone{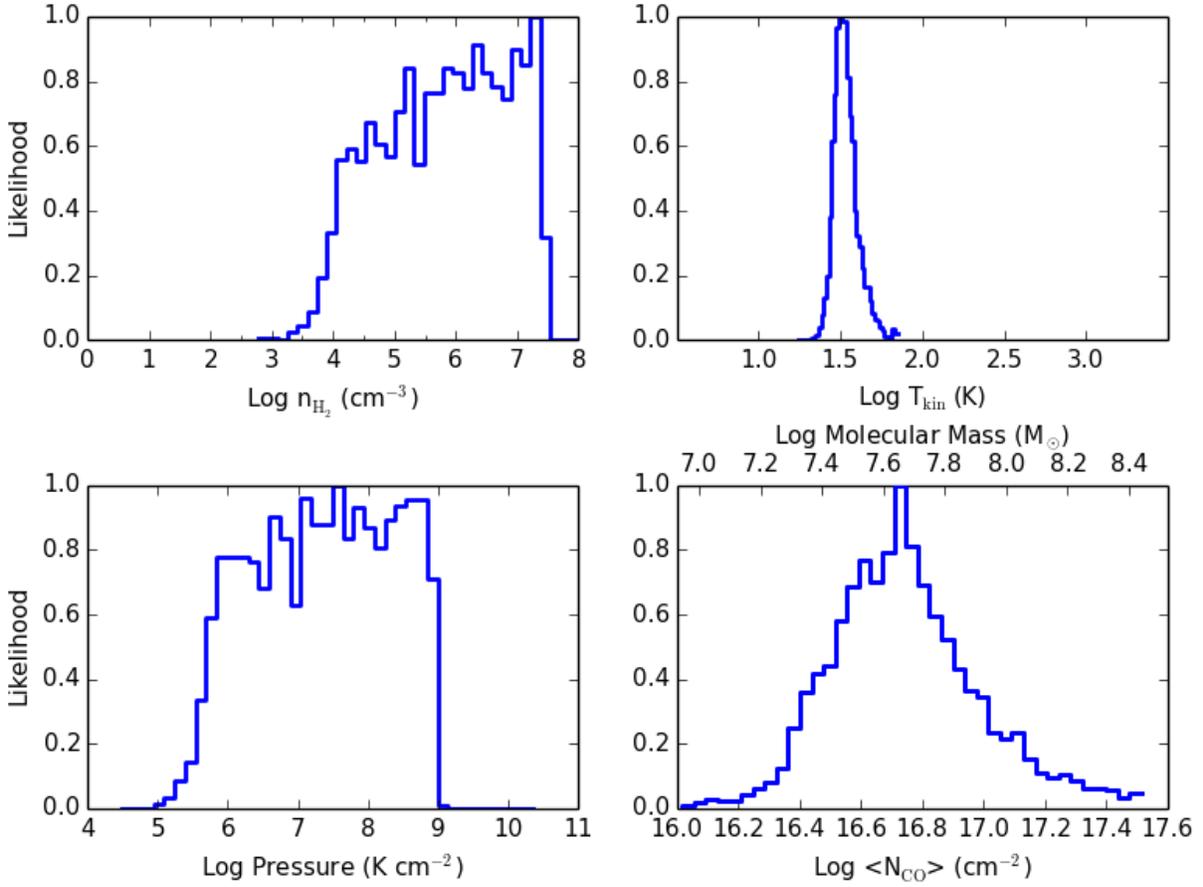}
\caption{Marginalized parameter likelihoods for the Broad component  using CO lines from $J = 1 - 0$ to $J = 8 - 7$.  Only one component is required to fit the Broad component spectral line energy distribution without an extinction correction or higher-J lines.  \label{fig6}} 
\end{figure}

\begin{figure}
\plotone{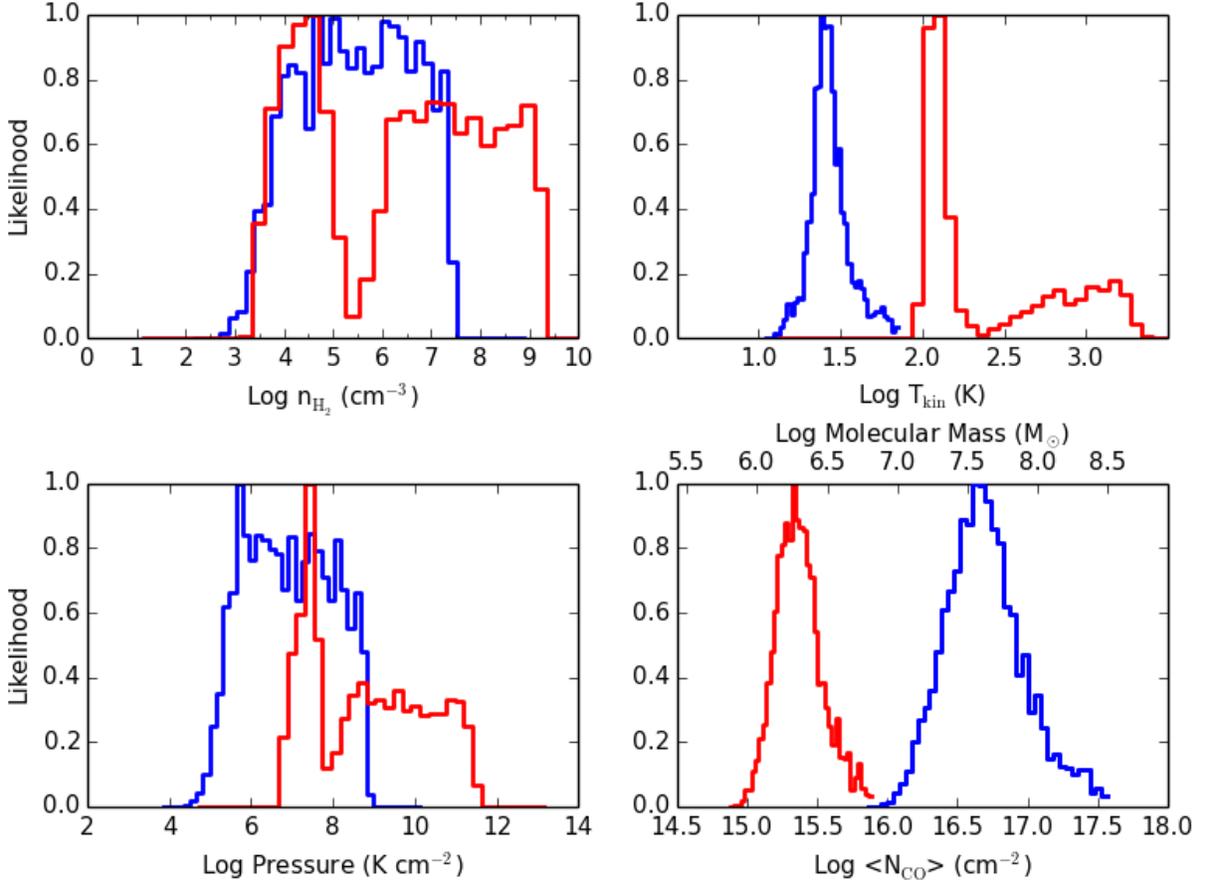}
\caption{Marginalized parameter likelihoods for the Broad component including the $J = 9 - 8$ to $J = 13 - 12$ lines, where the model CVC line fluxes are subtracted from the FTS line fluxes.  These models are subject to the constraint that $\mathrm{M_{warm} < M_{cold}}$.  The cold component is blue and the warm component is red. \label{fig7}} 
\end{figure}

\clearpage

\begin{figure}
\includegraphics[scale=1.0,angle=90]{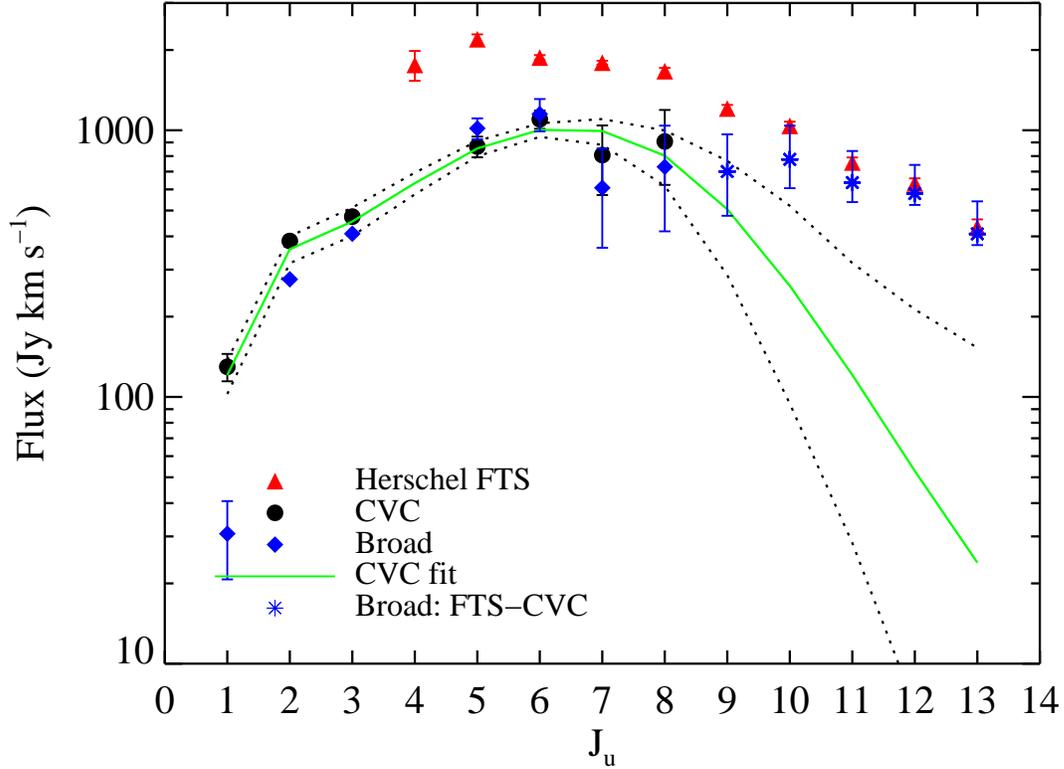}
\caption{(Non extinction-corrected) CO spectral line energy distribution, including {\it
    Herschel} FTS, {\it Herschel} HIFI, and ground-based observations
    from Alatalo et al. (2011). ``Broad" (blue diamonds) refers to the line fluxes from Gaussian profile fits.  The Broad component lines above $J = 8 - 7$ (``Broad: FTS$-$CVC", blue asterisks) were derived by subtracting the CVC fit (green line and black dotted $\pm 1\sigma$ lines) from the FTS line fluxes (red triangles), incorporating the line flux likelihoods to derive Broad component line flux error bars.
    All error bars are $1\sigma$.
  \label{fig8}} 
\end{figure}
\clearpage

\begin{figure}
\epsscale{0.9}
\plotone{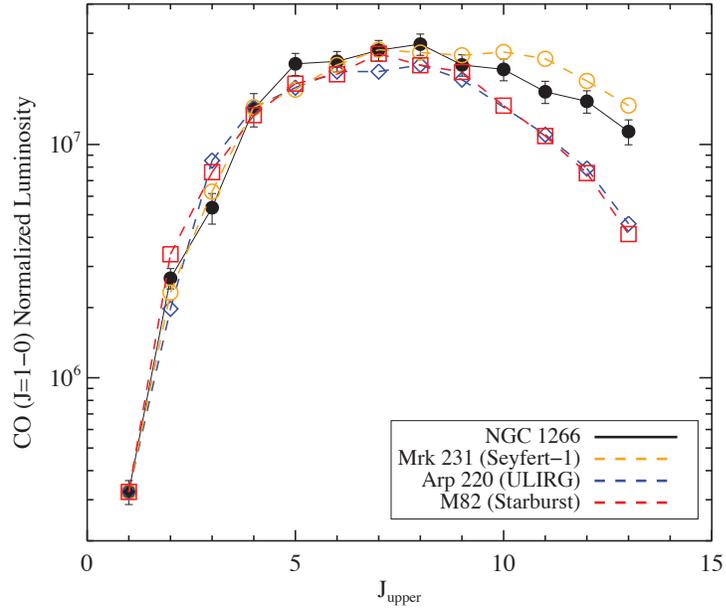}
\caption{A comparison of the CO spectral line energy distributions of
  Mk231 (Seyfert galaxy), NGC1266, and Arp 220 and M82, which are
  galaxies whose far-infrared luminosities are dominated by star
  formation.  The spectral line energy distributions are normalized at
  $J = 6 - 5$. 
  \label{fig9}}
\end{figure}

\clearpage

\begin{figure}
\epsscale{0.9}
\plotone{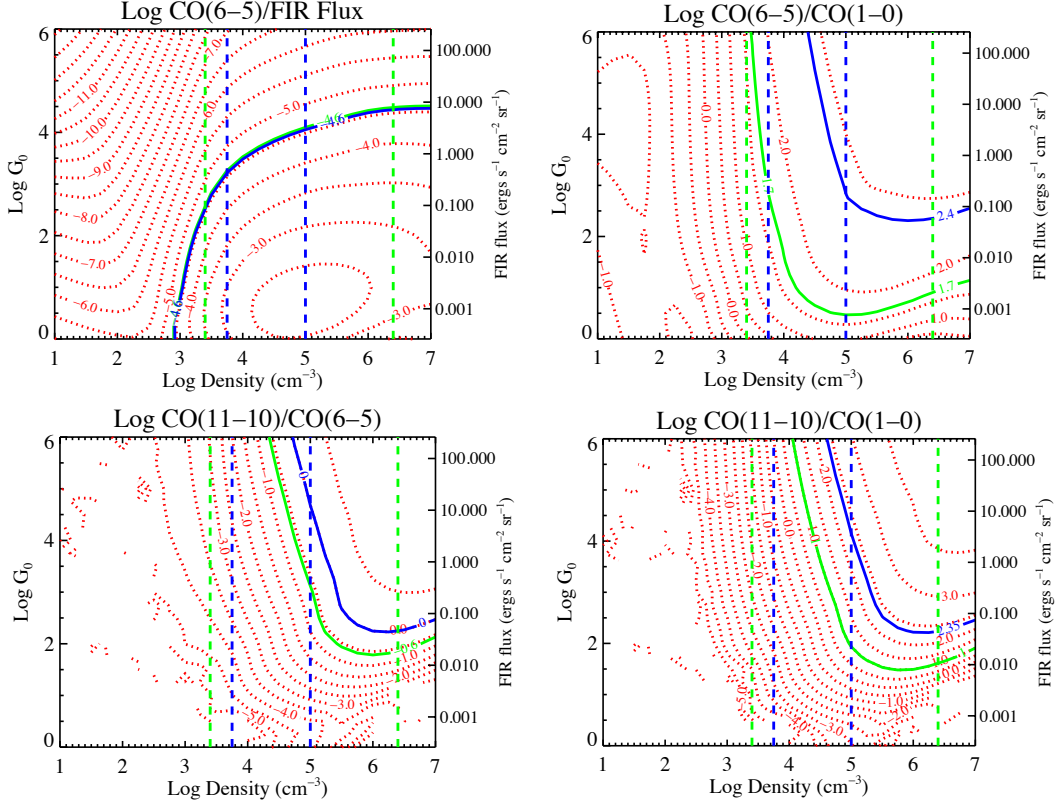}
\caption{PDR model grid, with gas density along the x-axis and UV
radiation intensity parameter $G_0$ on the y-axis, overplotted with
the observed ratios. The models are those of Wolfire et al (2010). (a)
The ratio of the CO $J = 6 - 5$ line flux to the far-infrared line
flux. The ratio observed for the CVC is in green, that for the Broad
component is in blue. The green (blue) vertical dashed lines show the
density ranges allowed by the likelihood models ($1\sigma$ for the CVC and $3\sigma$ for the Broad component). The observed
ratios have been calculated by assigning 100\% of the far-infrared
flux to each component separately; see text for discussion. The
right-hand vertical scale shows the model far-infrared surface
brightness corresponding to the values of $G_0$. (b) As in (a), but
for the $L_{\rm CO}(J = 6 - 5)/L_{\rm CO}(J = 1 - 0)$ line ratio. (c)
As in (a), but for the $L_{\rm CO}(J = 11 - 10)/L_{\rm CO}(J = 6 - 5)$
line ratio. Division of the $J = 11 - 10$ line into CVC and Broad
components is done using the best-fitting likelihood model; see
text. (d) As in (c), but for the $L_{\rm CO}(J = 11 - 10)/L_{\rm CO}(J
= 1 - 0)$ line ratio.
  \label{fig10}}
\end{figure}

\clearpage

\begin{deluxetable}{clcrc}
\tablecaption{{\it Herschel} Observations \label{tab1}}
\tablewidth{0pt}
\tabletypesize{\footnotesize}
\tablehead{\colhead{Observation}& \colhead{Instrument/Band} &
  \colhead{OBSID} & \colhead{OD} & \colhead{t$_\mathrm{integration}$} \\
    & & & & \colhead{(s)} 
}
\startdata
Lines \& Continuum & SPIRE-FTS & 1342239353 & 1012 & 5640 \\
CO $J = 5 - 4$ & HIFI 1b & 1342238647 & 996 & 4063 \\
CO $J = 6 - 5$ & HIFI 2a & 1342238179 & 985 & 3918 \\
CO $J = 7 - 6$ & HIFI 3a & 1342239585 & 1016 & 3945 \\
CO $J = 8 - 7$ & HIFI 3b & 1342237619 & 980 & 3928 \\
\enddata
\label{obs}
\end{deluxetable}

\clearpage

\begin{deluxetable}{ccccc}
\tablecaption{{\it Herschel} SPIRE-FTS Line Fluxes \label{tab3}}
\tablewidth{0pt}
\tabletypesize{\footnotesize}
\tablehead{
  \colhead{ID} & \colhead{Transition} & \colhead{Rest
    Frequency} & \colhead{Flux} & \colhead{$\sigma$} \\
  & & \colhead{(GHz)} & \colhead{(Jy km s$^{-1}$)} & \colhead{(Jy km
    s$^{-1}$)} 
}
\startdata
CI & $J = 1 - 0$ & 492.16064 & 789.4 & 135.3 \\
CI & $J = 2 - 1$ & 809.34198 & 1139.6	& 29.2 \\
$^{12}$CO & $J = 4 - 3$ & 461.04077 & 1757.2 & 225.4 \\
$^{12}$CO & $J = 5 - 4$ & 576.26794 & 2193.6 & 93.2 \\
$^{12}$CO & $J = 6 - 5$ & 691.47308 & 1873.3 & 40.1 \\
$^{12}$CO & $J = 7 - 6$ & 806.65179 & 1792.4 & 29.9 \\
$^{12}$CO & $J = 8 - 7$ & 921.79968 & 1665.4 & 48.0 \\
$^{12}$CO & $J = 9 - 8$ & 1036.91235 & 1206.0 & 38.4 \\
$^{12}$CO & $J = 10 - 9$ & 1151.98547 & 1038.6 & 38.2 \\
$^{12}$CO & $J = 11 - 10$ & 1267.01453 & 757.2 & 33.3 \\
$^{12}$CO & $J = 12 - 11$ & 1381.99512 & 631.7 & 28.6 \\
$^{12}$CO & $J = 13 - 12$ & 1496.92285 & 432.7 & 30.4 \\
NII & $^3P_1-^3P_0$ & 1461.1319 & 750.2 & 82.6 \\
H$_2$O & $2_{11}-2_{02}$ & 752.0332 & 430.4 & 28.9 \\
H$_2$O & $2_{02}-1_{11}$ & 987.92676 & 628.8 & 55.5 \\
H$_2$O & $3_{12}-3_{03}$ & 1097.36475 & 462.6 & 35.7 \\
H$_2$O & $3_{21}-3_{12}$ & 1162.91162 & 761.9 & 33.3 \\
H$_2$O & $4_{22}-4_{13}$ & 1207.63867 & 147.7 & 33.5 \\
H$_2$O & $2_{20}-2_{11}$ & 1228.78882 & 372.4 & 34.4 \\
H$_2$O & $5_{23}-5_{14}$ & 1410.61804 & 102.3 & 24.5 \\
               
\enddata
\label{obs1}
\end{deluxetable}

\clearpage

\begin{deluxetable}{lccc}
\tablecaption{CO Velocity Component Fluxes \label{tbl-4}}
\tablewidth{0pt}
\tablehead{
\colhead{Transition} & \colhead{CVC Flux}&
\colhead{Broad Component Flux} & \colhead{Broad Component Width} \\
\colhead{} & \colhead{(Jy km s$^{-1}$)} & \colhead{(Jy km s$^{-1}$)} & \colhead{(km s$^{-1}$)} }
\startdata
J $= 1 - 0$ & $130 \pm 3$ & $31 \pm 10$ & $597 \pm 108$ \\ 
J $= 2 - 1$ & $384 \pm 6$ & $276 \pm 4$ & $283 \pm 5$\\ 
J $= 3 - 2$ & $474 \pm 9$ & $409 \pm 6$ & $274 \pm 5$\\ 
J $= 5 - 4$ & $870 \pm 80$ & $1020 \pm 90$ & $328 \pm 35$\\ 
J $= 6 - 5$ & $1100 \pm 90$ & $1150 \pm 160$ & $273 \pm 68$\\ 
J $= 7 - 6$ & $810 \pm 240$ & $610 \pm 250$ & $234 \pm 126$\\ 
J $= 8 - 7$ & $910 \pm 280$ & $730 \pm 310$ & $205 \pm 145$\\ 
\enddata
\tablecomments{Uncertainties are $1\sigma$.  The CVC line widths were fixed at 123 km s$^{-1}$ (see text).}
\end{deluxetable}

\begin{deluxetable}{cccccc}
\tabletypesize{\footnotesize}
\tablecaption{Continuum Photometry \label{tab2}}
\tablewidth{0pt}
\tablehead{
\colhead{Wavelength} & \colhead{Flux Density} &
\colhead{$\sigma_{\mathrm{stat}}$} &
\colhead{$\sigma_{\mathrm{calib}}$} & \colhead{Facility} & Ref. \\
\colhead{($\mu$m)} & \colhead{(Jy)} & \colhead{(Jy)} &\colhead{(Jy)}
}
\startdata
 250 &     3.690 &     0.185 &     0.185 & SPIRE-PSW & 1 \\
 350 &     1.260 &     0.063 &     0.063 & SPIRE-PMW & 1 \\
 500 &     0.370 &     0.019 &     0.019 & SPIRE-PLW & 1 \\
& & & & & \\
  57 &    15.500 &     0.310 &     1.365 & {\it ISO} & 2 \\
  63 &    18.500 &     0.370 &     1.762 & {\it ISO} & 2 \\
  88 &    21.600 &     0.432 &     2.665 & {\it ISO} & 2 \\
 122 &    16.700 &     0.334 &     1.256 & {\it ISO} & 2 \\
 145 &    13.300 &     0.266 &     1.272 & {\it ISO} & 2 \\
 158 &    14.800 &     0.296 &     4.791 & {\it ISO} & 2 \\
 170 &    10.100 &     0.202 &     0.979 & {\it ISO} & 2 \\
  25 &     1.200 &     0.032 &     0.120 & {\it IRAS} & 3 \\
  60 &    13.130 &     0.045 &     1.313 & {\it IRAS} & 3 \\
 100 &    16.890 &     0.185 &     1.689 & {\it IRAS} & 3 \\
  24 &     0.843 &     0.008 &     0.034 & MIPS & 4, 5 \\
  70 &    10.304 &     0.118 &     0.515 & MIPS & 4, 5 \\
 160 &     6.339 &     0.144 &     0.761 & MIPS & 4, 5 \\
\enddata
\tablerefs{(1) This paper. (2) {\it ISO}:  Brauher et al. (2008). (3) {\it IRAS}:  Sanders et al. (2003). (4) and (5) Dale et al. (2007) and Temi et al. (2009), respectively.}
\end{deluxetable}

\clearpage

\begin{deluxetable}{cc}
\tablewidth{0pt}
\tablecaption{Dust Continuum Fit \label{tab4}}
\tabletypesize{\footnotesize}
\tablehead{
\colhead{Parameter} & \colhead{Mean $\pm1\sigma$}} 
\startdata
$T_{\mathrm{dust}}$ (K) & $56\pm3$ \\
$\lambda_0~(\mu$m$)$ & $206\pm25$ \\
$\beta$ & $2.00\pm0.29$ \\
$\alpha$ & $3.1\pm0.3$ \\
$\tau$~(100~$\mu$m) & $4.7\pm1.9$ \\
$\mathrm{Log}[L_{\mathrm{FIR}}~(\mathrm{L_{\odot}}\mathrm{,}~8-1000~\mu\mathrm{m})]$ & $10.44\pm0.01$  \\ 
$\mathrm{Log}[M_{\mathrm{dust}}$  ($\mathrm{M_{\odot}}$)] & $6.34\pm0.04$ \\ 
\enddata
\tablecomments{The dust mass uncertainty is statistical only:  it does not account for the uncertainty in the dust opacity, which could be a factor of two.}
\end{deluxetable}

\clearpage

\begin{deluxetable}{lcc}
\tablewidth{0pt}
\tablecaption{CVC Gas Parameters Constrained by Models}
\tabletypesize{\footnotesize}
\tablehead{
\colhead{Parameter} & \colhead{Cold Comp Median $\pm 1\sigma$} & \colhead{Warm Comp Median $\pm 1\sigma$}} 
\startdata
$T_{kin}$ (K) & $6^{+10}_{-2}$ & $74^{+130}_{-26}$ \\
Log$_{10}$($\langle N_{CO} \rangle_{\Omega}$) & $17.3^{+0.5}_{-0.4}$ & $16.3^{+0.5}_{-0.5}$ \\
Log$_{10}$(Mass) (M$_\odot$) & $8.3^{+0.5}_{-0.4}$ & $7.3^{+0.5}_{-0.5}$ \\
Log$_{10}$(Luminosity) (L$_\odot$ ) & $5.44^{+0.22}_{-0.18}$ & $6.90^{+0.16}_{-0.16}$ \\
\enddata
\end{deluxetable}

\clearpage

\begin{deluxetable}{lcc}
\tablewidth{0pt}
\tablecaption{Broad Component Gas Parameters Constrained by Models\label{Broad_all}}
\tabletypesize{\footnotesize}
\tablehead{
\colhead{Parameter} & \colhead{First Comp Median $\pm 1\sigma$} & \colhead{Second Comp Median $\pm 1\sigma$}} 
\startdata
$J = 8 - 7$ {\it and below} & & \\
$T_{kin}$ (K) & $36^{+5}_{-4}$ & ... \\
Log$_{10}$($\langle N_{CO} \rangle_{\Omega}$) & $16.7^{+0.2}_{-0.2}$ & ... \\
Log$_{10}$(Mass) (M$_\odot$) & $7.6^{+0.2}_{-0.2}$ & ... \\
Log$_{10}$(Luminosity) (L$_\odot$ ) & $6.84^{+0.06}_{-0.08}$ & ... \\
& & \\
{\it All lines} & & \\
$T_{kin}$ (K) & $27^{+9}_{-5}$ & $146^{+974}_{-27}$ \\
Log$_{10}$($\langle N_{CO} \rangle_{\Omega}$) & $16.7^{+0.3}_{-0.3}$ & $15.4^{+0.2}_{-0.1}$ \\
Log$_{10}$(Mass) (M$_\odot$) & $7.6^{+0.3}_{-0.3}$ & $6.3^{+0.2}_{-0.1}$ \\
Log$_{10}$(Luminosity) (L$_\odot$ ) & $6.49^{+0.11}_{-0.14}$ & $7.23^{+0.07}_{-0.04}$ \\

\enddata
\end{deluxetable}

\end{document}